\documentclass{JHEP3}

  \usepackage{latexsym,bm,amsmath,amssymb,amsfonts}
  \usepackage{epsfig,graphics,graphicx}
  \usepackage{slashed}

  \long\def\comment#1{ }

  \newcommand{\eqnum}[1]{Eq.~\eqref{#1}}

  \newcommand{\mcal}{\mathcal}
  \newcommand{\rme}{{\rm e}}
  \newcommand{\rmd}{{\rm d}}   %ELS%

  \newcommand{\Lam}{\Lambda_{{\rm QCD}}}

  \newcommand{\order}[1]{\mcal{O}{(#1)}}

  \newcommand{\beq}{\begin{eqnarray}}
  \newcommand{\eeq}{\end{eqnarray}}
  
 \def\simge{\mathrel{%
   \rlap{\raise 0.511ex \hbox{$>$}}{\lower 0.511ex \hbox{$\sim$}}}}
\def\simle{\mathrel{
   \rlap{\raise 0.511ex \hbox{$<$}}{\lower 0.511ex \hbox{$\sim$}}}}

\title{\rm \LARGE A lattice test of strong coupling behaviour
in QCD at finite temperature}

\author{E. Iancu\\Institut de Physique Th\'eorique de Saclay,
 F-91191 Gif-sur-Yvette, France\\
        E-mail: \email{Edmond.Iancu@cea.fr}}
\author{A. H.~Mueller\\Department of Physics, Columbia University, New York, NY
10027, USA\\
        E-mail: \email{amh@phys.columbia.edu}}

\abstract{We propose a set of lattice measurements which could test whether
the deconfined, quark--gluon plasma, phase of QCD shows strong coupling
aspects at temperatures a few times the critical temperature
for deconfinement, in the region where the conformal anomaly becomes
unimportant. The measurements refer to twist--two operators which are
not protected by symmetries and which in a strong--coupling scenario would
develop large, negative, anomalous dimensions, resulting in a strong
suppression of the respective lattice expectation values in the continuum
limit. Special emphasis is put on the respective operator with lowest spin
(the spin--2 operator orthogonal to the energy--momentum tensor within
the renormalization flow) and on the case of quenched QCD, where this
operator is known for arbitrary values of the coupling: this is the quark
energy--momentum tensor. The proposed lattice measurements could also test
whether the plasma constituents are pointlike (as expected at weak
coupling), or not.}

\begin{document}

\section{Introduction}
\label{Intro}

The heavy ion experiments at RHIC have given two rather surprising and
important results, namely the medium effects know as elliptic flow and
jet quenching turned out to be much larger than simple expectations based
on perturbative QCD. This has led to the picture that the deconfined
matter produced at RHIC is a nearly perfect fluid, so like a strongly
coupled plasma  (see, e.g., the review papers
\cite{Gyulassy:2004zy,Muller:2007rs} and references therein). The
coupling constant $\alpha_s=g^2/4\pi$ in QCD can never become large,
because of asymptotic freedom, but it can be of order one at scales of
order $\Lam$, and this might lead to an effectively strong--coupling
behaviour. It is notoriously difficult to do reliable estimates in QCD
when $\alpha_s\simeq 1$, so it has become common practice to look to the
strongly coupled ${\mathcal N}=4$ supersymmetric Yang--Mills (SYM) theory
for guidance as to general properties of strongly coupled field theories
at finite temperature (see the review papers
\cite{Son:2007vk,Iancu:2008sp,Gubser:2009sn} for details and more
references). Since conformal symmetry is an essential property of
${\mathcal N}=4$ SYM, this theory is probably not a good model for the
dynamics in QCD in the vicinity of the deconfinement phase transition,
where the conformal anomaly associated with the running of the coupling
in QCD is known to be important. But lattice studies \cite{Cheng:2007jq}
show that the relative conformal anomaly $(\epsilon -3p)/\epsilon$
($\epsilon$ is the energy density and $p$ is the pressure) decreases very
fast with increasing $T$ above $T_c$ and becomes unimportant (smaller
than 10\%) for temperatures $T\gtrsim 2 T_c\simeq 400$~MeV. Hence, there
is a hope that, within the intermediate range of temperatures at
$2T_c\lesssim T \lesssim 5T_c$, which is the relevant range for the
phenomenology of heavy ion collisions at RHIC and LHC, the dynamics in
QCD may be at least qualitatively understood by analogy with ${\mathcal
N}=4$ SYM theory at strong coupling.

One result suggested by SYM theory is that all the leading twist
operators that occur in the operator product expansion of deep inelastic
scattering, with the exception of the energy--momentum tensor, should
have vanishing expectation values in a strongly coupled plasma. For the
${\mathcal N}=4$ SYM plasma, whose strong--coupling limit can be studied
via the gauge/gravity duality
\cite{Maldacena:1997re,Gubser:1998bc,Witten:1998zw}, this result follows
from the fact that only protected operators --- those whose anomalous
dimension is zero because of a symmetry, or conservation law --- do not
acquire large negative anomalous dimensions
\cite{Maldacena:1997re,Gubser:1998bc,Witten:1998zw,Gubser:2002tv,Kotikov:2004er,Brower:2006ea}.
This result is in fact natural in any field theory whose coupling is
large. As one measures this theory at smaller and smaller space--time
scales, one uncovers more and more strong evolution {\em (branching)} of
the quanta of the theory \cite{Polchinski:2002jw,HIM1,HIM2,HIM3}. The
smallness of the higher dimensional operators in the leading--twist
series just represents the fact that the higher energy--moments of
``bare'' quanta are naturally small at strong coupling, because the
energy has been shared among many quanta via the branching process. At
finite temperature, it is natural to assume that the branchings have
taken place between the temperature scale $T$ and the ``hard'' resolution
scale $Q$, with $Q\gg T$, at which the operator is evaluated.

It is important to emphasize that the renormalization flow of the
operators and, in particular, their anomalous dimensions are determined
by the \emph{vacuum} properties of the theory --- the temperature enters
only as the natural scale at which this flow should begin (and which
therefore controls the early running of the coupling in a theory like
QCD). In particular, at weak coupling, the anomalous dimensions are
computable in the zero--temperature perturbative expansion, which is a
series in powers of $\alpha_s$. This should be contrasted with the
calculation of \emph{thermal} expectation values, so like the pressure,
whose perturbative expansion is quite subtle already at weak coupling,
and in particular non--analytic in $\alpha_s$, because of infrared
problems associated with the thermal Bose--Einstein distribution
\cite{Blaizot:2001nr,Blaizot:2003tw,Kraemmer:2003gd}. Thus a
non--perturbative study of the renormalization flow of the (unprotected)
leading twist operators would allow one to distinguish between genuine
strong--coupling effects and the failure of the perturbation theory due
to medium effects, which occurs already at weak coupling. This would
avoid the ambiguity inherent in the present lattice studies of the QCD
thermodynamics \cite{Cheng:2007jq}, whose results cannot be accommodated
by a strict perturbative expansion, yet they appear to be consistent, for
temperatures $T\gtrsim 2.5T_c$, with the predictions of the HTL--resummed
perturbation theory
\cite{Blaizot:1999ip,Blaizot:2000fc,Andersen:1999sf,Andersen:2002ey}, and
not too far away from the strong--coupling limit of ${\mathcal N}=4$ SYM
\cite{Gubser:1998nz,Blaizot:2006tk}.

In particular, recent lattice calculations \cite{Cheng:2008zh} of the
fluctuations of the electric charge, baryonic number, and strangeness in
the quark--gluon plasma appear to be remarkably close to the respective
results of HTL--resummed perturbation theory  \cite{Blaizot:2001vr} (and
also to the ideal gas limit) already for $T\gtrsim 1.5T_c$, thus strongly
supporting a quasiparticle picture of the weak coupling type. The fact
that the approach towards the ideal gas limit when increasing $T$ above
$T_c$ appears to be faster for the quark susceptibilities than for the
pressure or energy density, is perhaps to be attributed to the fact that
the conformal anomaly, which is so important for the thermodynamics in
the vicinity of $T_c$, is less important for fermionic observables, like
the above susceptibilities.

Now, there is \emph{a priori} no contradiction in having a quasiparticle
picture also at strong coupling, as shown by the fact that the entropy
density of the ${\mathcal N}=4$ SYM plasma in the strong--coupling limit
is close to the respective value at zero coupling. However if the
effective coupling is large, one expects the quasiparticles to be highly
composite, without a pointlike core carrying a significant fraction of
the quasiparticle energy. This is illustrated by recent calculations of
deep inelastic scattering (DIS) off the strongly coupled ${\mathcal N}=4$
SYM plasma, which show that, when this plasma is measured on a hard
resolution scale $Q\gg T$, one finds only low--energy constituents, with
energy fractions $x\lesssim T/Q\ll 1$ \cite{HIM2,HIM3}. The higher $Q$
is, the smaller are the energy fractions, meaning that there are no
pointlike constituents.

How to determine what is the corresponding picture for the quark--gluon
plasma ? Of course, one cannot literally perform a deep inelastic
scattering on the QCD matter produced at RHIC to test whether or not
there are pointlike constituents having energy of order of $T$ (in the
rest frame of the plasma). The only experimental evidence on this comes
from the phenomenology of ``jet quenching'', which within perturbative
QCD at least, is the process responsible for both energy loss and
transverse momentum broadening of a hard probe propagating through the
medium. The relevant transport coefficient $\hat q$ (the ``jet--quenching
parameter'') is given by \cite{Baier:1996kr}
 \beq \hat q\,=\,\frac{4\pi^2 \alpha_s N_c}{N_c^2-1}\,\frac{\rmd
 xG(x,Q^2)}{\rmd V}\,,\eeq
where $\rmd xG(x,Q^2)/\rmd V$ is the number of gluons per unit volume in
the plasma measured on the relevant energy ($x$) and virtuality ($Q^2$)
resolution scales. It is generally assumed that $Q$ is of the order of
the saturation momentum of the plasma, since this is the typical momentum
of the gluons exchanged between the jet and the medium. Weak--coupling
estimates of $\hat q$ using ideal gas formulas for the density of the
plasma constituents at the scale $T$ together with perturbative evolution
to the hard scale $Q$ give $\hat q\simeq (0.5\div1) {\rm GeV}^2/{\rm
fm}$, while phenomenology \cite{Abelev:2006db,Adare:2006nq} rather
suggests that $\hat q$ should be somehow larger, between 5 and 15 ${\rm
GeV}^2/{\rm fm}$. This difference supports the picture of strong
evolution in the plasma, and hence of strong coupling
\cite{CasalderreySolana:2007zz}. One should nevertheless keep in mind
that this phenomenology is quite difficult and not devoid of ambiguities:
strong assumptions are necessary in order to compute $\hat q$, and also
to extract its value from the RHIC data (see, e.g., the discussion in
\cite{Baier:2006fr}).

In view of the experimental difficulties, it is natural to ask whether
lattice gauge theory can illuminate this question. Computing the DIS
structure functions on the lattice is in principle possible: via the
operator product expansion and for sufficiently high--$Q^2$, the moments
of the structure functions can be related to expectation values of
operators with spin $n$ and (classical) dimension $n+2$ --- the leading
twist operators --- which form an infinite series (only the even values
of $n$ being relevant for DIS). Given the space--like kinematics of the
DIS process, these operator expectation values are effectively Euclidean,
and thus can be evaluated on the lattice. In order to reconstruct the
structure functions from their moments, one would need to measure a large
number (in principle, infinite) of the latter, which is practically
tedious, if not impossible. Indeed, operators with spin $n=4,6,8,...$
involve too many derivatives to be accurately evaluated in lattice QCD,
although some attempts were done in that sense, for the case of the
proton structure functions (see, e.g., \cite{Gockeler:2004wp}).

However, in order to answer the limited questions that we address here,
such a full reconstruction of the plasma structure functions is actually
not needed. What we instead propose is to measure the expectation value
of the unique leading--twist operator with $n=2$ which is not protected
by symmetries, and thus check whether the corresponding result is rapidly
vanishing when approaching the continuum limit --- as expected for a
strong--coupling dynamics ---, or rather it is slowly evolving away from
the respective ideal gas expectations --- as it should be the case at
weak coupling.

Specifically, consider the two leading--twist operators with $n=2$ in
QCD, that is
 \beq\label{Of}
 \mathcal{O}^{\mu\nu}_f\,\equiv\,\bar q\,\gamma^\mu iD^\nu
 q \,-\,\mbox{(trace)}\,,\eeq
and, respectively,
 \beq\label{Og}
 \mathcal{O}^{\mu\nu}_g\,\equiv\,-F^{\mu\alpha}_aF_{\alpha}^{\nu, a}
 \,+\,\frac{1}{4}\,g^{\mu\nu}F^{\alpha\beta}_aF_{\alpha\beta}^{a}\,.
 \eeq
(It is understood that the fermionic operator $\mathcal{O}^{\mu\nu}_f$
involves a sum over quark flavors and a symmetrization of the Lorentz
indices, and we neglect the masses of the quarks.) These two operators
are well defined only with a renormalization prescription, and thus
implicitly depend upon the resolution scale $Q^2$. Since they have the
same quantum numbers, they mix with each other under the renormalization
flow. The following linear combination yields the total energy--momentum
tensor,
 \beq\label{Tmn}
 T^{\mu\nu}\,=\,\mathcal{O}^{\mu\nu}_f\,+\,\mathcal{O}^{\mu\nu}_g\,,
 \eeq
which is a conserved quantity, and thus is insensitive to quantum
evolution (it does not depend upon $Q^2$). Clearly, this operator cannot
be used to test whether the plasma has pointlike constituents, or not.
Within perturbation theory, it is always possible to construct the linear
combination of $\mathcal{O}^{\mu\nu}_f$ and $\mathcal{O}^{\mu\nu}_g$
which is orthogonal to $T^{\mu\nu}$ within the renormalization flow and
therefore vanishes in the continuum limit $Q^2\to\infty$ (the respective
anomalous dimension being negative). This is the operator whose
expectation value we would like to measure on the lattice. But if the
coupling is strong, we do not know how to explicitly construct this
orthogonal combination.

Fortunately, there is a simpler version of the theory where the
identification of this operator becomes possible for any value of the
coupling: this is \emph{quenched} QCD. Loosely speaking, this is the
theory obtained from QCD after removing all the quark loops. On the
lattice, this is non--perturbatively defined by removing the fermionic
determinant from the QCD action. Note that the quark fields are still
present in this theory, but only as external probes. In particular, it
makes sense to evaluate the fermionic operator \eqref{Of} in quenched
QCD: at finite temperature, this amounts to computing the Matsubara Dirac
propagator in the background of the thermal fluctuations of the gauge
fields. Such a calculation effectively resums all the respective Feynman
graphs of QCD, except for those involving quark loops. For instance, if
the coupling is weak (\emph{i.e.}, for high enough temperatures) and for
a given resolution scale $Q^2$ which is not too hard, the expectation
value $\langle \mathcal{O}^{\mu\nu}_f(Q^2)\rangle_{T}$ should be close to
the respective value for an ideal fermionic gas, as given by the
Fermi--Dirac distribution. Moreover, when $Q\sim T$, the temperature $T$
is the only scale in the problem, so by dimensional arguments we expect
 \beq\langle
\mathcal{O}^{\mu\nu}_f(Q^2\sim T^2)\rangle_{T}\,\propto\,T^4
 \qquad\mbox{for any value of the coupling}\,.\eeq
(We implicitly assume here that the temperature is sufficiently high for
the QCD trace anomaly to become unimportant; in practice, $T\gtrsim
2T_c$.) On the other hand, in the continuum limit $Q^2\to \infty$ at
fixed $T$, the above expectation value must vanish:
 \beq\label{CONT}
 \langle\mathcal{O}^{\mu\nu}_f(Q^2\to
 \infty)\rangle_{T}\,\to\,0\qquad\mbox{(fixed $T$)}\,.\eeq
\eqnum{CONT} will be derived in Sect.~4, but it is easy to see how it
comes about. The quark can emit gluons --- the more so, the harder the
scale at which one probes its substructure. But the emitted gluons, as
well as those from the thermal bath, are not allowed to emit
quark--antiquark pairs. Hence, when the system is probed on a
sufficiently hard scale, most of the total energy appears in the gluon
fields. We thus see that, within quenched QCD, $\mathcal{O}^{\mu\nu}_f$
is the $n=2$ operator orthogonal to the (total) energy--momentum tensor.
In the continuum limit, the latter reduces to its gluonic component:
$T^{\mu\nu}\to \mathcal{O}^{\mu\nu}_g(Q^2)$ as $Q^2\to \infty$.

Whereas \eqnum{CONT} holds for any value of the coupling, the rapidity of
the evolution with increasing $Q^2$ --- \emph{i.e.}, the rate at which
$\langle\mathcal{O}^{\mu\nu}_f\rangle_{T}$ approaches to zero
--- depends upon the strength of the interactions. For a weak coupling,
this evolution would be quite slow; using lowest--order perturbative QCD,
we shall estimate in Sect.~4 that for a temperature $T\simeq 3T_c$ and an
inverse lattice spacing $a^{-1}\equiv Q\simeq 4$~GeV, the deviation of
$\langle\mathcal{O}^{00}_f\rangle_{T}$ from the corresponding ideal gas
value should not exceed 30\%. On the other hand, if the evolution is more
like at strong coupling, and if the measurements of $\hat q$ are
indicative of what should be expected in the strongly--coupled QCD
plasma, one would expect $\langle\mathcal{O}^{00}_f\rangle_{T}$ to be
reduced by a factor of 5 or more. Of course, all the conclusions that
could be drawn in this way would strictly apply to quenched QCD alone.
However, we expect real (unquenched) QCD to behave similarly (within the
same range of temperatures), because the asymptotic freedom property of
QCD is driven by gluon dynamics.

%As a matter of facts, without fermion loops, we expect the coupling to
%run faster.

\section{Leading--twist operators: from weak to strong coupling}

Although the main emphasis in this paper is not on the process of deep
inelastic scattering by itself, but rather on the lattice evaluation of
specific, low spin, leading--twist operators, it is nevertheless natural
to introduce these operators in the context of DIS and thus summarize
some of their properties to be used later on.

Within QCD, there are two infinite sequences of leading--twist operators:
the fermionic ones,
\beq\label{Onf}
 \mathcal{O}^{(n)\,\mu_1\cdots\mu_n}_f\,\equiv\,\bar q\,\gamma^{\{\mu_1}
 (iD^{\mu_2})\cdots (iD^{\mu_n\}})q
  \,-\,\mbox{(traces)}\,,\eeq
(the curly brackets around Minkowski indices mean symmetrization), and
the gluonic ones,
 \beq\label{Ong}
 \mathcal{O}^{(n)\,\mu_1\cdots\mu_n}_g\,\equiv\,-\,\frac{1}{2}\,
 F^{\{\mu_1\nu} (iD^{\mu_2})\cdots (iD^{\mu_{n-1}})F^{\mu_n\}}_{\quad\ \nu}
 \,-\,\mbox{(traces)}\,,\eeq
where a trace over color indices is implicit. Such operators have spin
$n$, classical dimension $d=n+2$, and hence twist $t=d-n=2$. For $n=2$,
we recover the operators expressing the energy--momentum tensor for
quarks and gluons, respectively, cf. Eqs.~\eqref{Of}--\eqref{Og}. For
even values of $n$, $n=2,4,6,\dots$, these operators enter the OPE of the
current--current correlator which determines the cross--section for the
standard DIS process, as mediated by the exchange of a space--like
photon. More precisely, if the expectation values of the operators are
evaluated directly at the resolution scale $Q^2$ for DIS (the virtuality
of the space--like photon), then the OPE involves only the quark
operators \eqref{Onf}. In practice, however, it is convenient to evaluate
the operators at some fixed renormalization scale $\mu$, or at some
intrinsic physical scale
--- say, the temperature for the case of DIS off the quark--gluon
plasma. In such a case, the operators at the DIS scale $Q^2$ are obtained
by following the renormalization flow from the original scale $\mu^2$ (or
$T^2$), and under this flow, the fermionic and gluonic operators having
the same spin mix with each other.

It is generally stated that the leading--twist operators dominate the OPE
for DIS for sufficiently high $Q^2$. This is strictly true only so long
as the coupling is not too strong, as shown by the example of ${\mathcal
N}=4$ SYM theory, where explicit calculations were also possible at
strong coupling. To illustrate this, consider the leading--twist
contributions to the moments of the DIS structure function $F_2(x,Q^2)$,
which can be expressed as (we follow the conventions in
Ref.~\cite{Peskin})
 \beq\label{Mn}
 \int_0^1\rmd x\,x^{n-2}\,F_2(x,Q^2) \,\simeq\,A_f^{(n)}(Q^2)\,,
 \eeq
where the approximate equality sign means that in the r.h.s. we have kept
only the twist--2 contribution. The quantity $A_f^{(n)}(Q^2)$ is the
expectation value of the spin--$n$ fermionic operator \eqref{Onf}
evaluated at the resolution scale $Q^2$ and with all the kinematical
factors (responsible for the Minkowski tensor structure and for the
actual dimension of the operator) stripped off. For instance, if $F_2$
refers to a proton with 4--momentum $P^\mu$, then
 \beq\label{avP}
 \langle P|\mathcal{O}^{(n)\,\mu_1\cdots\mu_n}_f(Q^2)|P\rangle\,=\,
 A_f^{(n)}(Q^2)\,2P^{\mu_1}\cdots P^{\mu_n}\,-\,\mbox{(traces)}\,,\eeq
whereas for a plasma at temperature $T$ :
 \beq\label{avT}
 \langle\mathcal{O}^{(n)\,\mu_1\cdots\mu_n}_f(Q^2)\rangle_T\,=\,
 A_f^{(n)}(Q^2)\,T^{n}\,
 2n^{\mu_1}\cdots n^{\mu_n}\,-\,\mbox{(traces)}\,,\eeq
where $n^\mu$ is the four--velocity of the plasma, with $n^\mu=(1,0,0,0)$
in the rest frame of the plasma. Note that $A_f^{(n)}$ and $F_2$ are
dimensionless in the case of the proton, but they have mass dimension two
in the case of the plasma; this difference is related to the
normalization of the proton wavefunction. The Bjorken $x$ variable is
defined as $x=Q^2/2(P\cdot q)$ for DIS off the proton and, respectively,
$x=Q^2/2T(n\cdot q)$ for DIS off the plasma; here, $q^\mu$ is the
4--momentum of the virtual photon, with $q^\mu q_\mu = -Q^2$.

Let us assume that the operator $\mathcal{O}^{(n)}_f$ is normalized at
the scale $\mu_0$ and ignore the issue of operator mixing for the time
being (we shall return to this issue in Sect.~3). The corresponding
operator at a different resolution scale $Q$ is obtained by solving the
renormalization group equation (henceforth the Minkowski indices will be
kept implicit)
 \beq\label{evolOn}
 \mu^2\frac{\rmd}{\rmd \mu^2} \ \mathcal{O}^{(n)}_f\,=\,
 \gamma_f^{(n)}\mathcal{O}^{(n)}_f\ \ \Longrightarrow\ \
 \mathcal{O}^{(n)}_f(Q^2)\,=\,\exp\Bigg\{ \int\limits_{\mu_0^2}^{Q^2}
 \frac{\rmd \mu^2}{\mu^2}\,\gamma_f^{(n)}(\mu^2)\Bigg\}\,
 \mathcal{O}^{(n)}_f(\mu_0^2)\,,
 \eeq
where $\gamma_f^{(n)}$ is the corresponding anomalous dimension, which in
QCD depends upon the scale because of the running of the coupling.
Clearly, a similar evolution equation applies for the expectation value
$A_f^{(n)}(Q^2)$ of the above operator. It turns out that the anomalous
dimensions are always negative, with the notable exception of the
energy--momentum tensor \eqref{Tmn}, for which $\gamma=0$. Hence,
$A_f^{(n)}(Q^2)\to 0$ as $Q^2\to \infty$ for any $n\ge 4$, whereas for
$n=2$ we have
 \beq\label{SR}
 \int_0^1\rmd x\,F_2(x,Q^2) \,\to\,{\rm const.}\quad\mbox{as}\quad
 Q^2\to \infty
 \,,
 \eeq
which is simply the statement of energy--momentum conservation. These
results imply that, when increasing $Q^2$, $F_2(x,Q^2)$ is increasing at
small $x$, but decreasing at large $x$: the evolution acts to decrease
the average value of the energy fraction of the partons in the
wavefunction. This should be expected given the physical picture of the
evolution in terms of parton branching, as described in the Introduction.
The rate of the evolution towards zero for the unprotected operators, and
also the weight of the small--$x$ partons in the sum--rule \eqref{SR},
are however quite different at weak and respectively strong coupling, as
we now explain.

Consider weak coupling first. To lowest order in perturbative QCD, the
anomalous dimensions are obtained as (for a generic leading--twist
operator $\mathcal{O}$)
 \beq
 \gamma_\mathcal{O}(\mu^2)\,=\,-a_\mathcal{O}\,\frac{\alpha_s(\mu^2)}{4\pi}
  \,=\,-\,\frac{a_\mathcal{O}}{b_0\ln(\mu^2/\Lam^2)}\,,\eeq
where $a_\mathcal{O}$ is a positive number and in writing the second
equality we have used the one--loop expression for the QCD running
coupling, with $b_0=(11N_c-2N_f)/3$. Then \eqnum{evolOn} implies
 \beq\label{EvolW}
 \mathcal{O}^{(n)}_f(Q^2)\,=\,\left[\frac{\ln(\mu_0^2/\Lam^2)}
 {\ln(Q^2/\Lam^2)}\right]^{{a^{(n)}_f}/{b_0}}\,
 \mathcal{O}^{(n)}_f(\mu_0^2)\,,
 \eeq
which shows that the approach towards zero with increasing $Q^2$ is
merely logarithmic. Still at weak coupling, consider the case of a
conformal field theory, so like ${\mathcal N}=4$ SYM, where the coupling
$\alpha=g^2/4\pi$ is fixed; then, $\gamma_\mathcal{O}=
-a_\mathcal{O}(\alpha/4\pi)$ is a fixed number of $\order{\alpha}$, and
 \beq\label{Evolconf}
 \mathcal{O}^{(n)}_f(Q^2)\,=\,\left[\frac{\mu_0^2}{Q^2}\right]^
 {a^{(n)}_f\frac{\alpha}{4\pi}}\,
 \mathcal{O}^{(n)}_f(\mu_0^2)\,,\eeq
so that the evolution is typically faster than in QCD, since it is not
slowed down by the decrease of the coupling with increasing $Q^2$. But
for both QCD and ${\mathcal N}=4$ SYM, the anomalous dimensions are small
$\sim\order{g^2}$ at weak coupling, so the leading--twist operators
dominate indeed the moments of the DIS structure functions at high $Q^2$
: the corresponding contributions from higher--twist operators are
suppressed by inverse powers of $Q^2$ with exponents of $\order{1}$.

Consider also the energy--momentum sum--rule \eqref{SR} : although
$F_2(x,Q^2)$ does rise at small $x$, as expected, pQCD predicts that this
rise is rather mild, so that the integral in \eqnum{SR} is dominated by
rather large values of $x$, of order 0.1. This is confirmed by the
experimental data at HERA, which can be parameterized by a law
$F_2(x,Q^2)\sim 1/x^{\lambda(Q^2)}$ where the effective exponent
$\lambda(Q^2)$ rises slowly with $Q^2$, but it remains relatively small:
$\lambda(Q^2)\simeq 0.15\div 0.3$. This expresses the fact that, at weak
coupling, the branching proceeds via bremsstrahlung and favors the
emission of small--$x$ gluons, whose number grows very fast, but which
carry only a tiny fraction of the energy of their parent partons.
Accordingly, most of the total energy remains in the ``valence'' degrees
of freedom at large $x$. Since this is true for arbitrarily high $Q^2$,
it is clear that these valence constituents can be viewed as pointlike.

What is the corresponding situation at strong coupling ? Since the
respective results are not known for QCD, we focus on the ${\mathcal
N}=4$ SYM theory, whose strong--coupling limit has been addressed via the
gauge/string duality. By ``strong coupling'', we more precisely mean here
the limit in which the gauge coupling is weak, $g^2\ll 1$, but the number
of colors is sufficiently large, $N_c\gg 1$, for the 't Hooft coupling to
be large: $\lambda\equiv g^2N_c \gg 1$. (Recall that the 't Hooft
coupling is the relevant coupling for perturbation theory at large
$N_c$.) Via the AdS/CFT correspondence, the (gluonic) leading--twist
operators are mapped onto excited string states --- closed strings which
rotate in the $AdS_5$ space--time geometry. By computing the energy
spectrum for such states, one can deduce the quantum dimensions
$\Delta(n)=n+2-2\gamma^{(n)}$ of the dual operators $\mathcal{O}^{(n)}$,
and thus extract their anomalous dimensions $\gamma^{(n)}$. One has thus
found \cite{Gubser:2002tv,Kotikov:2004er,Brower:2006ea}
 \beq\label{smalln}
 \gamma^{(n)}\,\simeq\,-\sqrt{\frac{n}{2}}\ \lambda^{1/4}
 \quad\mbox{for}\quad 1\,\ll\,n\,\ll\,\sqrt{\lambda}\,,
 \eeq
and, respectively,
 \beq\label{largen}
 \gamma^{(n)}\,\simeq\,-\frac{\sqrt{\lambda}}{2\pi}\,\ln\frac{n}
 {\sqrt{\lambda}}\quad\mbox{for}\quad n\,\gg\,\sqrt{\lambda}\,.\eeq
That is, the anomalous dimensions are again negative (except, of course,
for the protected energy--momentum tensor), and moreover they are
extremely large: of $\order{\lambda^{1/4}}$ for the operators with lower
spin. Via \eqnum{evolOn}, this implies that all the leading--twist
operators with the exception of $T^{\mu\nu}$ are strongly suppressed at
high $Q^2$, and hence they become irrelevant for DIS: the respective
structure functions are rather controlled by $T^{\mu\nu}$ together with
protected higher--twist operators which have zero anomalous dimensions.

The fact that the anomalous dimensions are so large at strong coupling
means that the branching process is very fast and, as a result of it, all
partons have fallen at small values of $x$. This is further confirmed by
the fact that the anomalous dimensions \eqref{smalln}--\eqref{largen}
rise with $n$, showing (via the moments \eqref{Mn}) that the support of
the structure function is now concentrated at small values of $x$.

As an illustration of the situation at strong coupling, let us recall the
results for DIS off the ${\mathcal N}=4$ SYM plasma in the
strong--coupling limit $\lambda\to\infty$ (or $N_c \to\infty$). In that
limit, the anomalous dimensions for the non--protected leading--twist
operators become infinite, while the higher--twist protected operators
cannot contribute to the DIS cross--sections because of the
energy--momentum conservation. Accordingly, the explicit calculation in
Ref. \cite{HIM2} finds that there is no power--like tail in $F_2(x,Q^2)$
at high $Q^2$. More interestingly, it also finds that there is an
exponential tail,
 \beq
    F_{2}(x,Q^2)\,\sim\,
    x{N^2_cQ^2}\,\exp\left\{-c\, \big(Q/Q_s(x)\big)\right\}
    \quad\mbox{for}  \quad Q\,\gg\,Q_s(x)\,=\,\frac{T}{x}
    %\,=\,\exp\big\{-c \big(x/x_s(Q)\big)\big\}
  \label{Ftunnel} \eeq
($c$ is a number), which reflects a tunneling process, reminiscent of the
Schwinger mechanism: the highly--virtual ($Q\,\gg\,Q_s(x)$) space--like
current can decay into charged partons via a tunnel effect induced by a
uniform force $\sim T^2$, which represents the action of the plasma on
the dipole fluctuations of the current in this large--$N_c$ limit.

The exponential in \eqnum{Ftunnel} can be alternatively rewritten as
$\exp\{-c(x/x_s(Q))\}$ with $x_s(Q)=T/Q$, showing that, for fixed $Q\gg
T$, the DIS structure function is essentially zero for any $x$ larger
than $x_s(Q)\ll 1$. This reflects the fact that, via successive
branchings, all partons have fallen at small values of $x$. And, indeed,
for sufficiently small values $x\lesssim x_s(Q)$ (or, equivalently, for
low enough virtualities $Q\lesssim Q_s(x)$ at a given $x$), the
exponential suppression goes away, and one finds \cite{HIM2}
 \beq
    F_{2}(x,Q^2)\,\sim\,
    x{N^2_cQ^2}\,\left(\frac{T}{xQ}\right)^{2/3}
    \quad\mbox{for}  \quad Q\,\lesssim\,Q_s(x)
    %\quad\mbox{(or $x\lesssim x_s(Q)$)}
    %\,=\,\exp\big\{-c \big(x/x_s(Q)\big)\big\}
  \,.\label{Flowx} \eeq
These estimates are such that the energy--momentum sum--rule \eqref{SR}
is saturated by the partons along the saturation line, \emph{i.e.}, those
having $x\simeq x_s(Q)$ :
 \beq T^2\int_0^1\rmd
 x\,F_2(x,Q^2)\,\simeq\,T^2\,x_sF_2(x_s,Q^2)\,\sim\, N_c^2 T^4
 \,. \label{SRT} \eeq
As also emphasized above, this sum--rule reproduces the right order of
magnitude for the energy density of the strongly--coupled plasma:
$\langle T^{00}\rangle_T\sim N_c^2 T^4$. One can similarly check that the
higher moments with $n\ge 4$ are power suppressed at high $Q^2$ :
   \beq\label{MnT}
 \int_0^1\rmd x\,x^{n-2}\,F_2(x,Q^2) \,\sim\,x_s^n N^2_cQ^2
 \,\sim\,N^2_cQ^2\left(\frac{T}{Q}\right)^n\,.
 \eeq

\section{Evolution of $n=2$ operators in QCD for a generic coupling}

In this section we describe the evolution of the $n=2$ leading--twist
operators. We focus on the respective flavor--singlet operators, of which
there are two: the quark ($\mathcal{O}^{\mu\nu}_f$) and gluon
($\mathcal{O}^{\mu\nu}_g$) energy--momentum tensors displayed in
Eqs.~\eqref{Of}--\eqref{Og}. Our emphasis will be on the mixing between
these two operators under quantum evolution, leading to two orthogonal
eigen--operators: one which is {\em a priori} known for any value of the
coupling, since this is protected by energy--momentum conservation and
hence it is scale--independent --- this is, of course, the total
energy--momentum tensor,
$T^{\mu\nu}=\mathcal{O}^{\mu\nu}_f+\mathcal{O}^{\mu\nu}_g$ ---, and the
other one which is not protected and hence it depends upon the
renormalization scale $Q^2$. The latter, that we shall denote as
$\Theta^{\mu\nu}(Q^2)$, is explicitly known in QCD only for sufficiently
high $Q^2$, where perturbation theory can be used to compute the matrix
of anomalous dimensions (see e.g. Ch. 18 in \cite{Peskin}). Here, we are
rather interested in the situation at generic, and relatively strong,
coupling, so our subsequent developments will be necessarily formal and
incomplete: we shall try and use physical constraints and guidance from
${\mathcal N}=4$ SYM theory in such a way to characterize the mixing
matrix and the structure of $\Theta^{\mu\nu}$ as well as we can without
performing explicit calculations in QCD.

In full generality, the relevant renormalization group equations can be
written in matrix form as (from now on we shall omit the Lorentz indices)
 \beq\label{Evoleq}
 \mu^2\frac{\rmd}{\rmd \mu^2}  \left(
\begin{array}{ccc} \mathcal{O}_g \\ \mathcal{O}_f\end{array}\right)
\,=\,
 \left(
\begin{array}{ccc} \gamma_{gg} & \gamma_{gf} \\
                   \gamma_{fg} & \gamma_{ff}
\end{array}
 \right)
 \left(
\begin{array}{ccc} \mathcal{O}_g \\ \mathcal{O}_f\end{array}\right),
 \eeq
which features the $2\times 2$ anomalous dimension matrix $\gamma(\mu^2)$
of the $n=2$ leading--twist operators. To lowest order in perturbation
theory, the scale dependence of $\gamma$, as encoded in the running
coupling $\alpha_s(\mu^2)$, factorizes out from the matrix structure. In
that case it is convenient to pursue the analysis by diagonalizing the
$\gamma$ matrix, since the corresponding eigenvectors are
scale--independent (see e.g. Ref.~\cite{Peskin}). However, such a
simplification does not occur for the general case at hand. It is then
preferable to consider the formal solution to \eqnum{Evoleq}, as obtained
by integrating this equation from the conventional renormalization scale
$\mu_0^2$ to the physically interesting scale $Q^2$. The solution reads
 \beq\label{EvolO}
 \left(
\begin{array}{ccc} \mathcal{O}_g(Q^2) \\ \mathcal{O}_f
 (Q^2)\end{array}\right)
\,=\,
 \left(
\begin{array}{ccc} M_{gg} & M_{gf} \\
                   M_{fg} & M_{ff}
\end{array}
 \right)
 \left(
 \begin{array}{ccc} \mathcal{O}_g(\mu_0^2)
 \\ \mathcal{O}_f(\mu_0^2)\end{array}\right),
 \eeq
where the evolution matrix $M=(M_{ij})$, with $i,\,j=g$ or $f$, can be
compactly written as
 \beq\label{M}
 M(Q^2,\mu_0^2)\,=\,{\rm P}\,\exp\Bigg\{ \int\limits_{\mu_0^2}^{Q^2}
 \frac{\rmd \mu^2}{\mu^2}\,\gamma(\mu^2)\Bigg\}\,,\eeq
where the symbol $P$ in the r.h.s. indicates the $\mu^2$--ordering of the
product of matrices in the series expansion of the exponential.

We shall now argue that, still in full generality, the matrix $M$ has
only two independent components. This follows from energy--momentum
conservation: the requirement that $T=\mathcal{O}_f+\mathcal{O}_g$ be
scale--independent, that is,
 \beq
 \mathcal{O}_g(Q^2)+\mathcal{O}_f(Q^2)\,=\,
 \mathcal{O}_g(\mu_0^2)+\mathcal{O}_f(\mu_0^2)\,,\eeq
implies two constraints on the components of the matrix $M$ (since the
operators at the original scale $\mu_0^2$ should be viewed as two
independent quantities), leading to
 \beq
 M_{fg}\,=\,1-M_{gg}\,,\qquad M_{gf}\,=\,1-M_{ff}\,,
 \eeq
and therefore\footnote{There are of course similar constraints on the
anomalous dimension matrix, which there imply $\gamma_{fg}=-\gamma_{gg}$
and $\gamma_{gf}=-\gamma_{ff}$. This means that, at any scale $\mu^2$,
the matrix $\gamma(\mu^2)$ has the left eigenvector $(1, 1)$ with
eigenvalue $\gamma_T=0$. But the other eigenvector, orthogonal to $T$, is
generally scale--dependent.}
 \beq\label{Mgen}
 M\,=\, \left(
\begin{array}{ccc} M_{gg} & 1- M_{ff} \\
                   1 - M_{gg} & M_{ff}
\end{array}
 \right).\eeq

So far we have not taken into account the fact that the coupling could be
strong. Recall that our objective is to give a test of the idea that QCD
is strongly coupled at a scale which is a few times the critical
temperature for deconfinement $T_c$. So, let us assume that the coupling
is strong at the scale $\mu_0$ at which one starts the evolution. (This
is the scale to be identified with the temperature $T$ when the evolution
takes place in the quark--gluon plasma phase.) Then we expect $M$ to have
an eigenvalue which is extremely small, nearly zero, corresponding to the
fact that the ``unprotected'' operator $\Theta$ has a large and negative
anomalous dimension, which is exponentiated by the evolution. Let us give
a more formal argument in that sense: to that aim, we divide the
logarithmic phase--space for the evolution $\ln(Q^2/\mu_0^2)$ into a
large number $N$ of small steps with width
$\epsilon=(1/N)\ln(Q^2/\mu_0^2)$, in such a way as to ensure that, within
each interval, the anomalous dimension matrix is essentially constant.
Then we can break the $\mu^2$--ordered exponential in \eqnum{M} into a
product of $N$ ordinary exponentials:
 \beq\label{Mprod}
 M(Q^2,\mu_0^2)\,=\,\rme^{\epsilon\gamma_N}\rme^{\epsilon\gamma_{N-1}}
 \cdots \rme^{\epsilon\gamma_1}\,,\eeq
where, of course, the quantities $\gamma_i\equiv\gamma(\mu^2_i)$ are
$2\times 2$ matrices. The determinant $\det M$ is equal to the product of
the determinants of the $N$ matrices in the r.h.s. For any such a matrix,
we can diagonalize $\gamma_i$ locally at $\mu_i^2$ : $\gamma_i = h_i
\,{\rm diag}(\gamma_T,\gamma_{\Theta(i)})\,h_i^{-1}$, where $\gamma_T=0$
(this is the anomalous dimension of the protected operator $T$), whereas
$\gamma_{\Theta(i)}$ is strictly negative (this is the anomalous
dimension of the unprotected operator $\Theta(\mu^2)$ at
$\mu^2=\mu_i^2$). Then, clearly
 \beq
 \det\, \rme^{\epsilon\gamma_i}\,=\, \rme^{\epsilon\gamma_{\Theta(i)}}\
 \Longrightarrow\ \det M\,=\,\exp\Bigg\{ \int\limits_{\mu_0^2}^{Q^2}
 \frac{\rmd \mu^2}{\mu^2}\,\gamma_\Theta(\mu^2)\Bigg\}\,,\eeq
where the integrand in the exponent is negative at any $\mu^2$. Now, let
us assume that for $\mu^2$ close to the lower limit $\mu_0^2$ (or
anywhere else along the way from $\mu_0^2$ to $Q^2$), the anomalous
dimension is extremely large, so like at strong coupling: this implies
that $\det M\approx 0$, as anticipated. By imposing $\det M=0$ in
\eqnum{Mgen}, one finds $M_{ff}=1 - M_{gg}$. We thus finally deduce the
following, particularly simple, expression for the evolution matrix (with
$m\equiv M_{gg}$)
 \beq\label{Mstrong}
 M\,=\, \left(
\begin{array}{ccc} m & m \\
                   1 - m & 1 - m
\end{array}
 \right)\,,\eeq
valid when the evolution takes place at least partially in a region in
$\mu^2$ where the coupling is strong. Using this form for $M$ in
\eqnum{EvolO}, one finds
 \beq\label{OgfS}
 \mathcal{O}_g(Q^2)\,=\,m\,T\,,\qquad
 \mathcal{O}_f(Q^2)\,=\,(1-m)\,T\,,\eeq
which allows us to identify the $n=2$ operator orthogonal to $T$,
\emph{i.e.}, the one which has evolved essentially down to zero on the
resolution scale $Q^2$ :
 \beq\label{Theta}
 \Theta(Q^2)\,\equiv\,(1-m)\mathcal{O}_g(Q^2)\,-\,m\mathcal{O}_f(Q^2)
 \,=\,0\,.\eeq
If the quantity $m=m(Q^2,\mu_0^2)$ were known theoretically, then
\eqnum{Theta} would be a prediction that could be tested in lattice gauge
theory. Unfortunately, we do not know how to determine this quantity
within the scenario that QCD is strongly coupled at low scales.

What we do know, however, is that $m$ should be independent of the
precise value of the scale $\mu_0^2$ at which one starts the evolution:
indeed, $m$ is rather determined by the largest value of $\mu^2$ at which
the anomalous dimension $\gamma_\Theta(\mu^2)$ is still large. To see
this, let us introduce an intermediate scale $\mu^2_1$, with
$\mu_0^2<\mu_1^2< Q^2$, and thus write $M=M_1M_2$, with
 \beq
 M_1={\rm P}\,\exp\Bigg\{ \int\limits_{\mu_0^2}^{\mu_1^2}
 \frac{\rmd \mu^2}{\mu^2}\,\gamma(\mu^2)\Bigg\}\qquad\mbox{and}\qquad
 M_2={\rm P}\,\exp\Bigg\{ \int\limits_{\mu_1^2}^{Q^2}
 \frac{\rmd \mu^2}{\mu^2}\,\gamma(\mu^2)\Bigg\}\,.\eeq
Now, assume that $\mu_1$ is such that the coupling is still strong in its
neighborhood, so that $M_1$ has the structure shown in \eqnum{Mstrong}
with $m\to m_1$. Then one can easily check that $M=M_1$, that is $m=m_1$,
and this even for a matrix $M_2$ which has the most general possible
structure, as shown in \eqnum{Mgen}. (But of course in QCD we would also
expect $M_2$ to be of the simpler form \eqref{Mstrong}, since if the
coupling is strong at some scale $\mu_1$, it is still strong at the
softer scale $\mu_0<\mu_1$.) Hence, if $\mu_S$ is the largest value at
which the coupling is still effectively strong, then we have
$m(Q^2,\mu_0^2)=m(Q^2,\mu_S^2)$ for any $\mu_0<\mu_S$.

Finally, one may worry that in QCD anomalous dimensions are scheme
dependent and that there is no meaning to say that $\gamma$ is large.
However, when $Q^2$ is large, the operators on the left hand side of
\eqnum{EvolO} have very little scheme dependence because $\alpha_s(Q^2)$
becomes small at large $Q^2$. The scheme dependence refers merely to the
ability to transfer contributions between the evolution matrix
$M(Q^2,\mu_0^2)$ and the operators $\mathcal{O}_g(\mu_0^2)$ and
$\mathcal{O}_f(\mu_0^2)$ at the original scale. If QCD behaves like a
strongly coupled field theory, then the operators $\mathcal{O}_g(Q^2)$
and $\mathcal{O}_f(Q^2)$ at the final scale are expressible in terms of
the (protected) energy--momentum tensor, as shown in \eqnum{OgfS}. We
have modeled our discussion to ressemble the situation in ${\mathcal
N}=4$ SYM theory (where there is no scheme dependence, because of the
conformal symmetry), but we recognize that in QCD one could choose
schemes in which a condition like \eqnum{Theta}
--- \emph{i.e.}, the vanishing of $\Theta$ at the scale $Q^2$ --- does
not follow from the evolution (\emph{i.e.}, from the particular structure
\eqref{Mstrong} of the evolution matrix $M$), but rather from the fact
that a relation between $\mathcal{O}_g$ and $\mathcal{O}_f$ similar to
\eqref{Theta} holds already at the original scale $\mu_0^2$.

\section{Evolution of $n=2$ operators in quenched QCD}

Because we are unable to specify a definite value for the quantity $m$ in
Eqs.~\eqref{Mstrong} and \eqref{Theta}, it is difficult to devise a test
of strong coupling behaviour in terms of $n=2$ leading--twist operators
using lattice gauge theory for full (unquenched) QCD. However, experience
with lattice calculations shows that there is generally not a large
difference between quenched and unquenched QCD. Thus if full QCD is
effectively a strongly coupled theory in the soft momentum region, one
would naturally expect the same to be true for quenched QCD. As mentioned
in the Introduction, quenched QCD consists in ignoring the quark loops,
so the matrix element $\gamma_{fg}$ of the anomalous dimension matrix in
\eqnum{Evoleq} must vanish. (Recall that this element describes a
transition from gluon to quark fields.) Since $\gamma_{gg}=-\gamma_{fg}$
by energy--momentum conservation, and similarly
$\gamma_{ff}=-\gamma_{gf}$, we deduce that the $\gamma$ matrix has a very
simple structure in quenched QCD:
 \beq\label{gamma}\gamma(\mu^2)\,=\, \left(
\begin{array}{ccc} 0\ & -\gamma_{ff} \\
                   0\ & \gamma_{ff}
\end{array}
 \right)
 \eeq
This structure is already telling us that the operator $\Theta$
orthogonal to the energy--momentum tensor $T=\mathcal{O}_f+\mathcal{O}_g$
is simply the quark operator $\mathcal{O}_f$. (Indeed, the matrix
\eqref{gamma} has the left eigenvector $(0, 1)$ with eigenvalue
$\gamma_{ff}< 0$.) It is furthermore clear that the rest of the
discussion of the renormalization group evolution for $n=2$ goes exactly
like in the previous section, so in particular Eqs.~\eqref{Mstrong} and
\eqref{Theta} are still true, but now with $m=1$ (since gluon fields
cannot change into fermions). Once again, \eqnum{Theta} with $m=1$
confirms that $\Theta=\mathcal{O}_f$. Thus, within quenched QCD, a
strong--coupling scenario predict $\mathcal{O}_f(Q^2)\simeq 0$ for a
sufficiently hard scale $Q^2$. At finite temperature, this in turn
implies that the average value of the energy carried by a {\em bare}
quark (one which is measured on a hard resolution scale $Q^2\gg T^2$)
which is in equilibrium with a strongly--coupled thermal bath of gluons
is very small,
 \beq\label{OfS}
 \langle\bar q\,\gamma_0 iD_0
 q\,(Q^2)\rangle_T \,\simeq\,0\,,\eeq
and in particular much smaller than the corresponding ideal--gas value
(the Stefan--Boltzmann law for a gas of free, massless, quarks):
 \beq\label{SB}
 \langle\bar q\,\gamma_0 iD_0
 q\rangle_T^{(0)} \,=\,N_f N_c\frac{7\pi^2}{60}\,T^4\,.\eeq

By contrast, in a weak coupling scenario, the corresponding lattice
result should be rather closed to the above ideal gas value, and slowly
departure from it with decreasing lattice spacing $a=1/Q$. One can easily
evaluate the leading order perturbative corrections to \eqref{SB}, and
thus get a better estimate for what should be the result at weak
coupling: using \cite{Peskin}
 \beq
 \gamma_{ff}\,=\,-
 -a_{ff}\,\frac{\alpha_s(\mu^2)}{4\pi}\,,
 \qquad a_{ff}\,=\,\frac{8}{3}\,C_F,\eeq
one finds (cf. \eqnum{EvolW} with $a^{(n)}_f\to a_{ff}$ and
$b_0=11N_c/3$)
 \beq\label{EvolOf}
\frac{\langle
\mathcal{O}_f(Q^2)\rangle}
 {\langle
\mathcal{O}_f(\mu_0^2)\rangle} \,=\,\left[\frac{\ln(\mu_0^2/\Lam^2)}
 {\ln(Q^2/\Lam^2)}\right]^{8C_F/3b_0}
 \,.
 \eeq
For example, for $Q=4$~GeV, $\Lam=200$~MeV, and $\mu_0=3T_c\simeq
600$~MeV, one finds that the perturbative evolution reduces the
ideal--gas result \eqref{SB} by about 30\%.

What could be the corresponding suppression in a strong--coupling
scenario ? It is of course very difficult to answer this question given
our impossibility to perform calculations in QCD at strong coupling. But
if the experimental results at RHIC for the jet quenching parameter $\hat
q$ \cite{Abelev:2006db,Adare:2006nq} --- which, we recall, suggest an
enhancement by roughly a factor of 5 with respect to the respective
weak--coupling estimate --- are indeed indicative of the strength of the
quantum evolution in the QCD plasma, then one might expect a similarly
strong reduction, by a factor of 5 or more, for the quark energy density
in quenched QCD. That such an expectation is not totally unreasonable
(within that strong--coupling scenario) can be also viewed via the
following argument:

Although there is no good reason to believe that the strong--coupling,
large--$n$, estimates for the anomalous dimensions in  ${\mathcal N}=4$
SYM theory, cf. \eqnum{smalln}, could be applied to the QCD problem at
hand, let us nevertheless do so, by lack of a better argument. Previous
studies in the literature, concerning the comparison between ${\mathcal
N}=4$ SYM and thermal QCD in the temperature range of interest, suggest
that a reasonable value for the QCD `t Hooft coupling to be used in this
context is $\lambda_{\rm QCD}\simeq 5.5$
\cite{Gubser:2006qh,Blaizot:2006tk}. (For instance, this is close to the
naive estimate $\lambda_{\rm QCD} =3g^2$, with the 2--loop QCD running
coupling $g^2(\bar\mu)$ evaluated at the scale $\bar\mu=2\pi T$.) Via
\eqnum{smalln}, this yields (for $n=2$) an anomalous dimension
$|\gamma_{ff}|\sim 1$. Assume now that there exists a window for
strong--coupling dynamics, within which $\mathcal{O}_f(\mu^2)$ evolves
according to \eqnum{Evolconf}. Then
 \beq\label{EvolOfconf}
\frac{\langle \mathcal{O}_f(\mu^2)\rangle}
 {\langle
 \mathcal{O}_f(\mu_0^2)\rangle} \,\sim\,\left(\frac{\mu_0^2}{\mu^2}
 \right)^{|\gamma_{ff}|}\,\sim\,\frac{\mu_0^2}{\mu^2}\,,\eeq
whereas the subsequent evolution from $\mu^2$ to the harder scale $Q^2$
takes place at weak coupling, and hence it is much slower, cf.
\eqnum{EvolOf}. Taking $\mu_0=3T_c\simeq 600$~MeV once again, it is clear
that a reduction by a factor of 5 or larger is achieved as soon as
$\mu\gtrsim 2\mu_0\sim 1.2$~GeV, that is, even if the strong--coupling
dynamics holds only in a rather narrow window. The current lattice QCD
results for the QCD pressure or energy density show a rather smooth
behaviour for temperatures $T > 3T_c$, with almost no variation from
$3T_c$ up to $6T_c$; hence, if it so happens that QCD is (effectively)
strongly--coupled at the scale $3T_c$, there is no reason why this should
not remain true until the slightly harder scale of $6T_c$.

To summarize, a lattice calculation for quenched QCD finding a  result
close to \eqref{SB} would show that the ``quasiparticles'' of quenched
QCD are close to being pointlike and that the theory is weakly coupled.
On the other hand, a much smaller result, cf. \eqref{OfS}, would be
compelling evidence for an effectively strongly--coupled theory, with
quasiparticles (if they exist) highly composite.

\subsection*{Acknowledgments}

We would like to thank Sourendu Gupta, Frithjof Karsch, Andr\'e Morel,
and Bengt Petersson for useful discussions at various stages of this
work, and Yuri Dokshitzer, Keijo Kajantie and Tony Rebhan for a careful
reading and useful comments on the manuscript. A.H.~M. wishes to thank
the IPhT Saclay and the CPHT Ecole Polytechnique for their hospitality
and support while this work has been done. The work of E.~I. is supported
in part by Agence Nationale de la Recherche via the programme
ANR-06-BLAN-0285-01. The work of A.H.~M. is supported in part by the US
Department of Energy.

\providecommand{\href}[2]{#2}\begingroup\raggedright\endgroup

\end{document}